\documentclass[prb,twocolumn,showpacs,eqsecnum,preprintnumbers]{revtex4}
\usepackage{amssymb}
\usepackage{amsmath}
\usepackage{graphicx}
\usepackage{dcolumn}
\usepackage{bm}

\def\p{\partial}

\def\be{\begin{equation}}
\def\ee{\end{equation}}
\def\bea{\begin{eqnarray}}
\def\eea{\end{eqnarray}}
\def\bse{\begin{subequations}}
\def\ese{\end{subequations}}

\def\Hcone{H_{\text{c}1}}
\def\Hctwo{H_{\text{c}2}}

\draft
\begin{document}

\title{Skyrmion versus vortex flux lattices in $p\,$-wave superconductors}
\author{Qi Li$^{1}$, John Toner$^{1}$, and D. Belitz$^{1,2}$}
\affiliation{$^{1}$Department of Physics and Institute of Theoretical
Science, University
of Oregon, Eugene, OR 97403\\
$^{2}$Materials Science Institute, University of Oregon, Eugene, OR 97403 }
\date{\today}

\begin{abstract}
$p$-wave superconductors allow for topological defects known as skyrmions, in
addition to the usual vortices that are possible in both $s$-wave and $p$-wave
materials. In strongly type-II superconductors in a magnetic field, a skyrmion
flux lattice yields a lower free energy than the Abrikosov flux lattice of
vortices, and should thus be realized in $p$-wave superconductors. We
analytically calculate the energy per skyrmion, which agrees very well with
numerical results. From this, we obtain the magnetic induction $B$ as a
function of the external magnetic field $H$, and the elastic constants of the
skyrmion lattice, near the lower critical field $\Hcone$. Together with the
Lindemann criterion, these results suffice to predict the melting curve of the
skyrmion lattice. We find a striking difference in the melting curves of vortex
lattices and skyrmion lattices: while the former is separated at all
temperatures from the Meissner phase by a vortex liquid phase, the skyrmion
lattice phase shares a direct boundary with the Meissner phase. That is,
skyrmions lattices {\it never} melt near $\Hcone$, while vortex lattices {\it
always} melt sufficiently close to $\Hcone$. This allows for a very simple test
for the existence of a skyrmion lattice. Possible $\mu$SR experiments to detect
skyrmion lattices are also discussed.
\end{abstract}

\pacs{74.50.+r,74.70.Pq,74.25.Fy}
\maketitle


\section{Introduction}
\label{sec:I}

\begin{figure}[t,h]
\includegraphics[width=5cm]{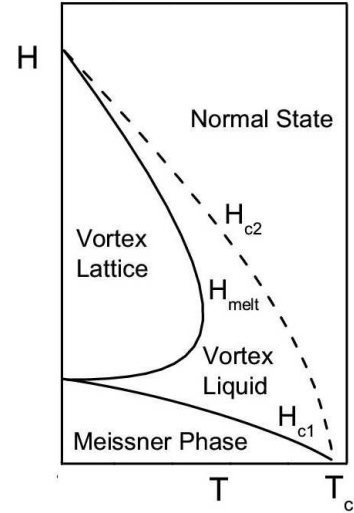}
\vskip -0mm
\caption{External field ($H$) vs. temperature ($T$) phase diagram for vortex
  flux lattices. Shown are the Meissner phase, the vortex lattice phase, the
  vortex liquid, and the normal state. Notice that the vortex lattice is never
  stable sufficiently close to $\Hcone$.}
\label{fig:1}
\end{figure}

One of the most fascinating phenomena exhibited by conventional, $s$-wave,
type-II superconductors is the appearance of an Abrikosov flux lattice of
vortices in the presence of an external magnetic field ${\bm H}$ in a range
$\Hcone < \vert{\bm H}\vert < \Hctwo$ between a lower critical field $\Hcone$
and an upper critical field $\Hctwo$.\cite{Tinkham_1975} It has been known for
quite some time both theoretically\cite{Huberman_Doniach_1979, Fisher_1980,
Nelson_Seung_1989, Brandt_1989} and experimentally \cite{Gammel_et_al_1988,
Safar_et_al_1992} that these flux lattices can melt. The melting curve
separates an Abrikosov vortex lattice phase from a vortex liquid phase, and the
vortex lattice is found to melt in the vicinity of both $\Hcone$ and $\Hctwo$,
as shown in Fig.\ \ref{fig:1}. The melting occurs because the elastic constants
of the flux lattice (i.e., the shear, bulk, and tilt moduli) vanish
exponentially near these field values. As a result, in clean superconductors,
root-mean-square positional thermal fluctuations $\sqrt{\langle \vert{\bm
u}({\bm x})\vert^2\rangle}$ grow exponentially as these fields are approached.
According to the Lindemann criterion, when these fluctuations become comparable
to the lattice constant $a$, the translational order of the flux lattice is
destroyed; i.e., the lattice melts.

\begin{figure}[t,h]
\vskip -1mm
\includegraphics[width=8.5cm]{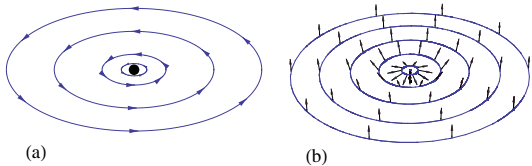}
\vskip -0mm
\caption{Order parameter configurations showing a vortex (a), and a skyrmion
(b). The local order parameters are represented by arrows on loci of equal
distance from the center of the defect. If the order parameter space is
two-dimensional, only vortices are possible, and there is a singularity at the
center of each vortex, (a). If the order-parameter space is three-dimensional,
a skyrmion can form instead, where the spin direction changes smoothly from
``down'' at the center to ``up'' at infinity, (b).}
\label{fig:2}
\end{figure}

Vortices are topological defects in the texture of the superconducting order
parameter, and in $s$-wave superconductors, where the order parameter is a
complex scalar, only one type of defect is possible. In $p$-wave
superconductors, the more complicated structure of the order parameter allows
for an additional type of topological defect known as a skyrmion. In contrast
to vortices, skyrmions do not involve a singularity at the core of the defect;
rather, the order parameter field is smooth everywhere, as illustrated in Fig.\
\ref{fig:2}.
Skyrmions were first introduced in a nuclear physics context by
Skyrme,\cite{Skyrme_1961} and slight variations of this
concept\cite{skyrmion_footnote} were later shown or proposed to be important in
superfluid $^3$He,\cite{Anderson_Toulouse_1977, Salomaa_Volovik_1987} in the
Blue Phases of liquid crystals,\cite{Wright_Mermin_1989} in Quantum Hall
systems,\cite{Sondhi_et_al_1993, Timm_Girvin_Fertig_1998} in itinerant
ferromagnets,\cite{Roessler_Bogdanov_Pfleiderer_2006} and in $p$-wave
superconductors.\cite{Knigavko_Rosenstein_Chen_1999} In the latter case,
skyrmions carry a quantized magnetic flux, as do vortices, although the lowest
energy skyrmion contains two flux quanta, while the lowest energy vortex
contains just one. For strongly type-II superconductors, skyrmions have a lower
free energy than vortices, and a vortex lattice should thus be the state that
occurs naturally.\cite{Knigavko_Rosenstein_Chen_1999}

Recent evidence of $p$-wave superconductivity in
Sr$_2$RuO$_4$\cite{Nelson_et_al_2004, Rice_2004, SrRuO_footnote} provides a
motivation for further exploring the properties of skyrmion flux lattices in
such systems.\cite{Knigavko_footnote} It was shown numerically by Knigavko et
al.\cite{Knigavko_Rosenstein_Chen_1999} that the interaction between skyrmions
falls off only as $1/R$ with distance $R$, as opposed to the exponentially
decaying interaction between vortices. As result, skyrmion lattices have a very
different dependence of the magnetic induction on the external magnetic field
near $\Hcone$ than do vortex lattices. In this paper we confirm and expand on
these results. We show analytically that the skyrmion-skyrmion interaction, in
addition to a leading $1/R$-dependence, has a correction proportional to $\ln
R/R^2$ that explains a small discrepancy between the numerical results in Ref.\
\onlinecite{Knigavko_Rosenstein_Chen_1999} and a strict $1/R$ fit, and we
calculate the interaction energy up to $O(1/R^2)$. We further show that the
melting curve of a skyrmion lattice is qualitatively different from that of a
vortex lattice. Namely, skyrmion lattices melt {\it nowhere} in the vicinity of
$\Hcone$, so there is a direct transition from the Meissner phase to the
skyrmion lattice, see Fig.\ \ref{fig:8} below. Finally, we predict and discuss
the magnetic induction distribution $n(B)$ of a skyrmion lattice state as
observed in a muon spin resonance ($\mu$SR) experiment. For a vortex lattice,
the exponential decay of the magnetic induction $B$ at large distances from a
vortex core implies $n(B) \propto \ln B/B$. For a skyrmion lattice, we find
that $B$ decays only algebraically, which leads to $n(B) \propto B^{-3/2}$.
Some of these results have been reported before in Ref.\
\onlinecite{Li_Toner_Belitz_2007}.

The paper is organized as follows. In Sec.\ \ref{sec:II} we review the
formulation in Ref. \onlinecite{Knigavko_Rosenstein_Chen_1999} of the skyrmion
problem. In particular, we start from the Ginzburg-Landau (GL) model for
$p$-wave superconductors and consider the free energy in a London
approximation. We parameterize the skyrmion solution of the saddle-point
equations, and express the energy in terms of the solution of the saddle-point
equations. In Sec.\ \ref{sec:III} we analytically solve these saddle-point
equations perturbatively for large skyrmion radius $R$, and we calculate the
energy of a single skyrmion as a power series in $1/R$ to order $1/R^2$. In
Sec.\ \ref{sec:IV} we determine the elastic properties of the skyrmion lattice,
and we predict the magnetic induction distribution $n(B)$ as observed in a
$\mu$SR experiment.

\section{Formulation of the skyrmion problem}
\label{sec:II}

In this section we review the formulation of the  skyrmion problem presented in
Ref.\ \onlinecite{Knigavko_Rosenstein_Chen_1999}, who  derived an effective
action that allows for skyrmions as saddle-point solutions. The resulting
ordinary differential equations (ODEs) describing skyrmions
\cite{Knigavko_Rosenstein_Chen_1999} are the starting point for our analytic
treatment.

\subsection{The action in the London approximation}
\label{subsec:II.A}

We start from a Landau-Ginzburg-Wilson (LGW) functional appropriate for
describing spin-triplet superconducting order,
\bse
\label{eqs:2.1}
\be
S = \int d{\bm x}\ {\cal L}({\bm\psi}({\bm x}),{\bm A}({\bm x})),
\label{eq:2.1a}
\ee
with an action density
\bea
{\cal L}({\bm\psi},{\bm A}) &=& t\vert{\bm\psi}\vert^2
                                 + u\,\vert{\bm\psi}\vert^4
                                 + v\vert{\bm\psi}\times{\bm\psi}^{\ast}\vert^2
                                 + \frac{1}{2m}\vert{\bm D\,\psi}\vert^2
\nonumber\\
&&  + \frac{1}{8\pi}\, ({\bm\nabla}\times{\bm A})^2.
\label{eq:2.1}
\eea
\ese
Here ${\bm\psi}({\bm x})$ is a 3-component complex order parameter
field,\cite{OP_footnote} ${\bm A}({\bm x})$ is the electromagnetic vector
potential, and ${\bm D} = {\bm\nabla }-iq{\bm A}$ denotes the gauge invariant
gradient operator. $m$ and $q$ are the mass and the charge, respectively, of a
Cooper pair, and we use units such that $\hbar = c = 1$. $t$, $u$, and $v$ are
the parameters of the LGW theory.

Let us look for saddle-point solutions to this action. In a large part of
parameter space, namely, for $v<0$ and $u>-v$, the stable saddle-point solution
is has the form ${\bm\psi}({\bm x}) \equiv {\bm\psi} = f_0\,(1,i,0)/\sqrt{2}$,
where the amplitude $f_0$ is determined by minimization of the free
energy.\cite{Vollhardt_Woelfle_1990} This is known as the $\beta$-phase, and it
is considered the most likely case to be realized in any of the candidates for
$p$-wave superconductivity.\cite{beta_phase_footnote} Fluctuations about this
saddle point are conveniently parameterized by writing the order parameter
field as
\bea
{\bm\psi}({\bm x}) = \frac{1}{\sqrt{2}}\,f({\bm x})\left(\hat{\bm n}({\bm x}) +
i\hat{\bm m}({\bm x})\right),
\label{eq:2.2}
\eea
where $\hat{\bm n}({\bm x})$ and $\hat{\bm m}({\bm x})$ are unit real
orthogonal vectors in order-parameter space and $f({\bm x})$ is the modulus of
order parameter. With this parameterization, the action density can  be written
\bea
{\cal L} &=& t\,f^2 + (u + v) f^4
\nonumber\\
&& + \frac{1}{2m}\Bigl[({\bm \nabla} f)^{2} + f^2
[\frac{1}{2}(\partial_i\hat{\bm l})^2 + (\hat{\bm n}\cdot\partial_i
\hat{\bm m} - q A_i)^2]\Bigr] \nonumber\\
&& + \frac{1}{8\pi}({\bm \nabla} \times {\bm A})^2,
\label{eq:2.3}
\eea
where $\hat{\bm l} = \hat{\bm n}\times \hat{\bm m}$, summation over repeated
indices is implied, and we have made use of the identities listed in Appendix
\ref{app:A}.

There are two length scales associated with the action density, Eq.\
(\ref{eq:2.3}).  The coherence length $\xi$ is determined by comparing
the $f^2$ term with the $({\bm\nabla}f)^2$ term,
\bse
\label{eqs:2.4}
\bea
\xi = 1/\sqrt{2m\vert t\vert}.
\label{eq:2.4a}
\eea
It is the length scale over which the amplitude of the order parameter will
typically vary. The London penetration depth $\lambda$ is determined by
comparing the ${\bm A}^2$ term  with the $({\bm\nabla}\times{\bm A})^2$
term,
\bea
\lambda = \sqrt{m/4\pi q^2\langle f \rangle^2}.
\label{eq:2.4b}
\eea
\ese
The ratio of these two length scales, $\kappa \equiv \lambda/\xi$, is the
Ginzburg-Landau parameter. Now we write $f({\bm x}) = f_0 + \delta f({\bm x})$,
with $f_0 = \sqrt{-t/2(u+v)}$. Deep inside the superconducting phase, where
$-t>0$ is large, the amplitude fluctuations $\delta f$ are massive, and to
study low-energy excitations one can integrate out $f$ in a tree approximation.
This approximation becomes exact in the limit of large $\kappa$ and is known in
this context as the London approximation. We introduce dimensionless quantities
by measuring distances in units of $\lambda$ and the action in units of
$\Phi_0^2/32\pi^3\lambda$, and we introduce a dimensionless vector potential
${\bm a} = 2\pi\lambda {\bm A}/\Phi_0$, with $\Phi_0 = 2\pi/q$ the magnetic
flux quantum. Ignoring constant contributions to the action we can then write
the action density in London approximation as
follows,\cite{Knigavko_Rosenstein_Chen_1999}
\bea
{\cal L}_{\text{L}} = \frac{1}{2}\,(\p_i{\hat{\bm l}})^2 + ({\hat{\bm
n}}\p_i{\hat{\bm m}} - a_i)^2 + {\bm b}^2,
\label{eq:2.5}
\eea
with ${\bm b} = {\bm\nabla}\times{\bm a}$. The above derivation makes it clear
that this effective action is a generalization of the $O(3)$ nonlinear sigma
model (represented by the first term on the right-hand side of Eq.\
(\ref{eq:2.5})) that one obtains for a real $3$-vector order parameter by
integrating out the amplitude fluctuations in tree
approximation.\cite{Zinn-Justin_1996}

\subsection{Saddle-point solutions of the effective action}
\label{subsec:II.B}

We now are looking for saddle-point solutions to the effective field theory,
Eq.\ (\ref{eq:2.5}). Considering ${\hat{\bm l}}$ and ${\hat{\bm n}}$
independent variables, and minimizing with respect to ${\hat{\bm l}}$ subject
to the constraints ${\hat{\bm l}}^2 = {\hat{\bm n}}^2 = 1$ and ${\hat{\bm
l}}\cdot{\hat{\bm n}} = 0$ yields
\bse
\label{eqs:2.6}
\bea
{\bm\nabla}^2 {\hat{\bm l}} - {\hat{\bm l}}({\hat{\bm
l}}\cdot{\bm\nabla}^2{\hat{\bm l}}) + 2 J_i ({\hat{\bm l}}\times \p_i{\hat{\bm
l}}) = 0,
\label{eq:2.6a}
\eea
with
\bea
{\bm J} = {\bm\nabla}\times {\bm b}
\label{eq:2.6b}
\eea
the supercurrent. The variation with respect to ${\bm a}$ is straightforward
and yields a generalized London equation,
\bea
a_i + J_i = {\hat{\bm n}} \p_i {\hat{\bm m}}.
\label{eq:2.6c}
\eea
\ese
It is convenient to take the curl of Eq.\ (\ref{eq:2.6c}) and use Eq.\
(\ref{eq:A.3}) to express the right-hand side of the resulting equation in
terms of ${\hat{\bm l}}$. We then obtain the saddle-point equations as a set of
coupled partial differential equations (PDEs) in terms of ${\bm b}$ and
${\hat{\bm l}}$ only:
\bse
\label{eqs:2.7}
\bea
b_i - {\bm\nabla}^2 b_i = \frac{1}{2}\,\epsilon_{ijk}\,{\hat{\bm l}}\cdot(\p_j
{\hat{\bm l}} \times \p_k {\hat{\bm l}}),
\label{eq:2.7a}\\
{\bm\nabla}^2 {\hat{\bm l}} - {\hat{\bm l}}({\hat{\bm
l}}\cdot{\bm\nabla}^2{\hat{\bm l}}) + 2 \epsilon_{ijk}\,\p_j b_k ({\hat{\bm
l}}\times \p_i{\hat{\bm l}}) = 0.
\label{eq:2.7b}
\eea
\ese
Notice that the right-hand side of Eq.\ (\ref{eq:2.7a}) is valid in  this
form only at points where ${\hat{\bm l}}({\bm x})$ is differentiable, see
Eq.\ (\ref{eq:A.3}). Field configurations that obey these PDEs have an
energy
\bea
E = \int d{\bm x}\ \left[ \frac{1}{2}(\p_i{\hat{\bm l}})^2 + ({\hat{\bm
n}}\cdot \p_i{\hat{\bm m}} - {\bm a})^2 + {\bm b}^2 - 2{\bm h}\cdot{\bm
b}\right]. \nonumber\\
\label{eq:2.8}
\eea
where we have added a uniform external magnetic field ${\bm h}$ measured in
units of $\Phi_0/2\pi\lambda^2$. Notice that the energy depends on ${\hat{\bm
n}}$ and ${\hat{\bm m}}$, whereas Eqs.\ (\ref{eqs:2.7}) depend only on
${\hat{\bm l}}$, and that different choices of ${\hat{\bm n}}$ and ${\hat{\bm
m}}$ can lead to the same ${\hat{\bm l}}$. Therefore, a field configuration
satisfying Eqs.\ (\ref{eqs:2.7}) is only necessary for making the energy
stationary, but not sufficient.

\subsubsection{Meissner solution}
\label{subsubsec:II.B.1}

\begin{figure}[t,h,b]
\includegraphics[width=8.5cm]{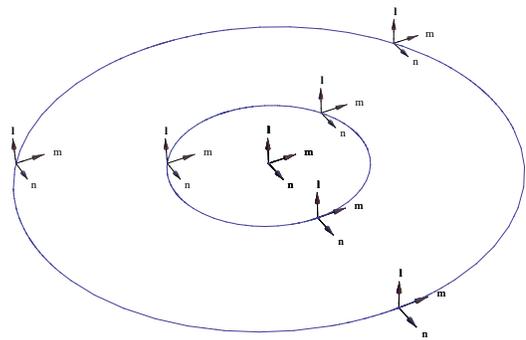}
\vskip -0mm
\caption{Configurations of the vectors $\hat{\ell}$, $\hat{m}$, and $\hat{n}$
  in a Meissner phase. All three vectors point in the same direction
everywhere.}
\label{fig:3}
\end{figure}

A very simple order parameter configuration consists of constant ${\hat{\bm
n}}({\bm x})$ and ${\hat{\bm m}}({\bm x})$ everywhere, see Fig.\ \ref{fig:3}.
This leads to an ${\hat{\bm l}}({\bm x}) \equiv {\hat{\bm l}}$  that is
constant everywhere.  Equation (\ref{eq:2.7b}) is then trivially satisfied. The
right-hand side of Eq. (\ref{eq:2.7a}) vanishes, and hence the PDE for ${\bm
b}$ reduces to the usual London equation with a solution ${\bm b}({\bm x})
\equiv 0$ in the bulk. This solution describes a Meissner phase with energy
$E_{\text{M}} = 0$.

\subsubsection{Vortex solution}
\label{subsubsec:II.B.2}

\begin{figure}[t,h,b]
\includegraphics[width=8.5cm]{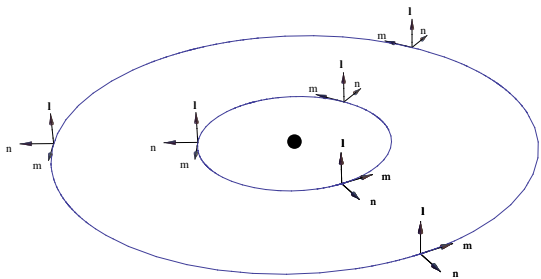}
\vskip -0mm
\caption{Configurations of the vectors $\hat{\ell}$, $\hat{m}$, and $\hat{n}$
  for a vortex. ${\hat\ell}$ is constant, whereas $\hat{m}$ and $\hat{n}$ rotate
  about the vortex core. Notice that the vector shown in Fig.\
 \ref{fig:2}(a) is $\hat{n}$.}
\label{fig:4}
\end{figure}

Now consider a field configuration where ${\hat{\bm n}}({\bm x})$ and
${\hat{\bm m}}({\bm x})$ are confined to a plane  (say, the $x$-$y$
plane), but rotate about an arbitrarily chosen point of origin:
\bea
{\hat{\bm n}}({\bm x}) &=& (\cos\phi,\sin\phi,0), \nonumber\\
{\hat{\bm m}}({\bm x}) &=& (-\sin\phi,\cos\phi,0),
\label{eq:2.9}
\eea
where $\phi$ denotes the azimuthal angle in the $x$-$y$ plane with respect to
the $x$-axis. This field configuration, known as a vortex and shown in Fig.
\ref{fig:4}, corresponds to a constant ${\hat{\bm l}}$ everywhere except at the
origin, where there is a singularity.
Therefore, the
right-hand side of Eq.\ (\ref{eq:2.7a}) is not applicable, and
we return to Eq.\ (\ref{eq:2.6c}), which takes the form
\bea
a_i + \epsilon_{ijk} \p_j a_k = \p_i \phi.
\label{eq:2.10}
\eea
For any closed path ${\cal C}$ in the $x$-$y$ plane that surrounds the origin
one has
\bse
\label{eqs:2.11}
\bea
\oint_{\cal C} d{\bm\ell}\cdot{\bm\nabla}\phi({\bm x}) = 2\pi,
\label{eq:2.11a}
\eea
or, by Stokes' theorem,
\bea
\int_{\cal A} d{\bm s}\cdot({\nabla} \times {\bm\nabla}\phi({\bm x})) = 2\pi,
\label{eq:2.11b}
\eea
\ese
where ${\cal A}$ is the surface whose boundary is ${\cal
C}$.\cite{n=1_footnote} This quantization condition  shows that, instead
of Eq.\ (\ref{eq:2.7a}), we have
\bea
{\bm b}({\bm x}) - {\bm\nabla}^2 {\bm b}({\bm x}) = 2\pi {\hat
z}\,\delta(x)\,\delta(y).
\label{eq:2.12}
\eea
This is solved by a ${\bm b}$ that is equal to the boundary condition value
everywhere along the $z$-axis and that falls off exponentially away from the
$z$-axis. This solution is known as a vortex, and the amount of magnetic flux
contained in one vortex is one flux quantum $\Phi_0$.\cite{n=1_footnote} It is
the precise analog of, and, indeed, essentially identical to, the familiar
vortex in conventional s-wave superconductors.

The energy of a vortex given by Eq.\ (\ref{eq:2.12}), as calculated from Eq.\
(\ref{eq:2.8}), is logarithmically infinite. This is due to the point-like
nature of the vortex core where the amplitude of the order parameter goes
discontinuously to zero. In reality, the amplitude cannot vary on length scales
shorter then the coherence length $\xi$, which provides an ultraviolet cutoff.
The energy is then proportional to $\ln\kappa$.\cite{Tinkham_1975} In an
external magnetic field this energy cost is offset by the magnetic energy gain
due to letting some flux penetrate the sample. For $\kappa$ larger than a
critical value $\kappa_c = 1/\sqrt{2}$, and for external fields larger than the
lower critical field $\Hcone$, a hexagonal lattice of vortices has a lower
energy than the Meissner phase. This state is known as an Abrikosov flux
lattice,

and is precisely the same as that in conventional s-wave
superconductors.\cite{Tinkham_1975}

\subsubsection{Skyrmion solution}
\label{subsubsec:II.B.3}

Due to the three-component nature of the order parameter, more complicated
solutions of the saddle-point equations can be constructed for which the vector
${\hat{\bm l}}$ is not fixed. Let $\theta$ be the angle between ${\hat{\bm l}}$
and the $z$-axis, and consider a cylindrically symmetric field configuration
parameterized as
\bea
{\hat{\bm l}} &=&{\hat{\bm e}}_z \cos\theta(r) + {\hat{\bm e}}_r\sin\theta(r),
\nonumber\\
{\hat{\bm n}} &=&\left({\hat{\bm e}}_z\sin\theta(r) - {\hat{\bm
e}}_r\cos\theta(r)
    \right)\sin\varphi + {\hat{\bm e}}_{\varphi}\cos\varphi
\nonumber\\
{\hat{\bm m}} &=&\left({\hat{\bm e}}_z\sin\theta(r) - {\hat{\bm
e}}_r\cos\theta(r)
    \right)\,\cos\varphi - {\hat{\bm e}}_{\varphi}\sin\varphi.
\nonumber\\
\label{eq:2.13}
\eea
For this to minimize the energy, ${\hat{\bm l}}$ at large distances from the
origin must be constant because of the first term in the energy, Eq.\
(\ref{eq:2.8}), and for a skyrmion centered in a cylinder of radius $R$ we take
${\hat{\bm l}}$ to point in the $+z$-direction for $r=R$, $\theta(r=R)=0$. The
quantization condition analogous to Eq.\ (\ref{eq:2.11b}) for the vortex
is\cite{Rajaraman_1982, Q=1_footnote}
\be
\int dx\,dy\ \epsilon_{ij}\,{\hat{\bm l}} \cdot (\p_{i}{\hat{\bm l}}\times \p_j
{\hat{\bm l}}) = 8\pi
\label{eq:2.14}
\ee
To be consistent with this, ${\hat{\bm l}}$ must point in the $-z$-direction at
the origin, $\theta(r=0)=\pi$.

Equation (\ref{eq:2.13}) parameterizes the order parameter in terms of a
function $\theta(r)$. In addition, the energy depends on the vector potential
which we take to be purely azimuthal, in accordance with our cylindrically
symmetric {\it ansatz},
\be
{\bm a}({\bm x}) = a(r)\,{\hat{\bm e}_{\varphi}}.
\label{eq:2.15}
\ee

With this parameterization, we obtain from Eq.\ (\ref{eq:2.8}) the energy per
unit length, along the cylinder axis, of a cylindrically symmetric skyrmion in
a region of radius $R$,
\bea
E/E_0 &=& \frac{1}{2} \int_0^R dr\,r\left[\left(\theta'(r)\right)^2
        + \frac{1}{r^2}\,\sin^2 \theta(r)\right]
\nonumber\\
&&+ \int_0^R dr\,r\left[\frac{1}{r}\,\left(1 + \cos\theta(r)\right) +
a(r)\right]^2
\nonumber\\
&&+ \int_0^R dr\,r \left[\frac{a(r)}{r} + a'(r)\right]^2,
\label{eq:2.16}
\eea
where $E_0 = (\Phi_0/4\pi\lambda)^2$. This expression was first obtained in
Ref.\ \onlinecite{Knigavko_Rosenstein_Chen_1999}. The three terms correspond to
the three terms in the London action, Eq.\ (\ref{eq:2.5}). They represent the
energy of the nonlinear sigma model, the kinetic energy of the supercurrent,
and the magnetic energy, respectively. Minimization of $E$ with respect to
$\theta(r)$ and $a(r)$ yields Euler-Lagrange equations
\bse
\label{eqs:2.17}
\bea
\theta''(r) + \frac{1}{r}\,\theta'(r) =
\frac{-\sin\theta(r)}{r}\,\left[\frac{2+\cos\theta(r)}{r} + 2 a(r)\right],
\nonumber\\
\label{eq:2.17a}\\
a''(r) + \frac{1}{r}\,a'(r) - \frac{1}{r^2}\,a(r) = a(r) + \frac{1}{r}\,\left[1
+ \cos\theta(r)\right].
\nonumber\\
\label{eq:2.17b}
\eea
\ese
This set of coupled, nonlinear ODEs must be solved subject to the boundary
conditions $\theta(r=0) = \pi$ and $\theta(r=R) = 0$, as explained above. The
solution is known as a skyrmion, and each skyrmion contains two flux
quanta.\cite{Q=1_footnote} Since Eqs.\ (\ref{eqs:2.7}) are necessary for making
the energy stationary, the solution of Eqs.\ (\ref{eqs:2.17}), inserted in
Eqs.\ (\ref{eq:2.13}, {\ref{eq:2.15}), is guaranteed to be a solution of Eqs.\
(\ref{eqs:2.7}) as well.

The energy of a single skyrmion is finite even in London approximation, see
Sec.\ \ref{sec:III} below. For large values of the Ginzburg-Landau parameter
$\kappa$ a skyrmion therefore has a lower energy than a vortex, and the value
of the lower critical field $\Hcone$, at which the Meissner phase becomes
unstable, is correspondingly lower for skyrmions than for vortices. This is the
basis for the expectation that, in strongly type-II (i.e., large-$\kappa$)
$p$-wave superconductors, a skyrmion flux lattice will be realized rather than
a vortex flux lattice.

\section{Analytic solution of the single-skyrmion problem}
\label{sec:III}

We now need to solve the coupled ODEs (\ref{eqs:2.17}). Due to their nonlinear
nature, this is a difficult task, and in Ref.\
\onlinecite{Knigavko_Rosenstein_Chen_1999} it was done numerically. It turns
out, however, that one can construct a perturbative analytical solution in the
limit of large skyrmion radius, $R \gg \lambda$, with $\lambda/R$ as a small
parameter. This provides information about the superconducting state near
$\Hcone$, where the system is always in that limit. We will construct the
perturbative solution, and calculate the energy, to second order in the small
parameter. Our general strategy is as follows. We use Eq.\ (\ref{eq:2.17b}) to
iteratively express $a$ in terms of $\theta$ and its derivatives. Substitution
in Eq.\ (\ref{eq:2.17a}) then yields a closed ODE for $\theta(r)$ that has to
be solved.

\subsection{Zeroth order solution}
\label{subsec:III.A}

Let us first consider $R=\infty$. For $r \to \infty$, the left-hand side of
Eq.\ (\ref{eq:2.17b}) falls off as $1/r^2$, and hence the vector potential, to
zeroth order for large $r$, is given by
\bea
a_{\infty}(r) = -\frac{1}{r}\left[1 + \cos\theta(r)\right].
\label{eq:3.1}
\eea
Note that we use the {\em exact} $\theta(r)$ in this expression, {\em not} the
zeroth order approximation to it. Since we can only  compute $\theta(r)$
perturbatively, this expression for the zeroth order vector potential will
itself have to be expanded perturbatively later. Substitution in Eq.\
(\ref{eq:2.17a}) yields
\bea
r^2\,\theta''(r) + r\,\theta'(r) = \frac{1}{2}\,\sin(2\theta(r)).
\label{eq:3.2}
\eea
The solution obeying the appropriate boundary condition
is\cite{Knigavko_Rosenstein_Chen_1999}
\bse
\label{eqs:3.3}
\bea
\theta_{\infty}(r) = f(r/\ell),
\label{eq:3.3a}
\eea
with
\bea
f(x) = 2\arctan(1/x).
\label{eq:3.3b}
\eea
\ese
The length scale $\ell$ is arbitrary at this point and will be determined later
from the requirement $\theta(r=R<\infty)=0$. For $R\gg 1$ it will turn out that
$\ell \propto \sqrt{R}$. The skyrmion solution is schematically shown in Fig.\
\ref{fig:5}.
\begin{figure}[t,h,b]
\includegraphics[width=8.5cm]{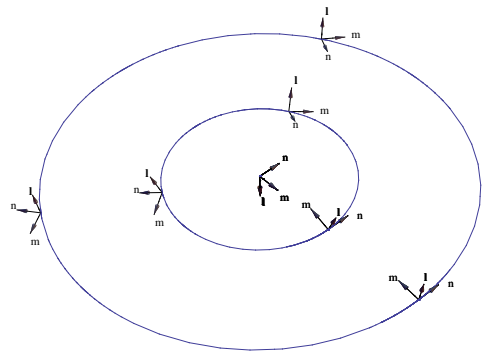}
\vskip -0mm
\caption{Configurations of the vectors $\hat{\ell}$, $\hat{m}$, and $\hat{n}$
  for a skyrmion. Notice that the vector shown in Fig.\ \ref{fig:2}(b) is
  $\hat{\ell}$.}
\label{fig:5}
\end{figure}

\subsection{Perturbation theory for $R\gg 1$}
\label{subsec:III.B}

We now determine the corrections to the zeroth order solution. Let us write
$\theta(r) = \theta_{\infty}(r) + \delta\theta(r)$ and $a(r) = a_{\infty}(r) +
\delta a(r)$ and require $\vert\delta a(r)\vert \ll \vert a_{\infty}(r)\vert$
and $\vert\delta\theta(r)\vert \ll 1$.\cite{expansion_footnote} An inspection
of the ODEs (\ref{eqs:2.17}) shows that for $r\alt O(\ell)$ , the corrections
can be expanded in a series in powers of $1/\ell$,
\bse
\label{eqs:3.4}
\bea
\delta\theta(r) &=& \frac{1}{\ell^2}\,g(r/\ell) + \frac{1}{\ell^4}\,h(r/\ell) +
O(1/\ell^6),
\label{eq:3.4a}\\
\delta a(r) &=& \frac{1}{\ell^3}\,\alpha(r/\ell) +
\frac{1}{\ell^5}\,\beta(r/\ell) + O(1/\ell^7).
\label{eq:3.4b}
\eea
\ese
The functions $\alpha$ and $\beta$ can be determined by substituting Eq.\
(\ref{eq:3.4b}) in Eq.\ (\ref{eq:2.17b}) and equating coefficients of powers of
${1/\ell}$. The resulting equations for $\alpha$ and $\beta$ are linear {\it
algebraic} equations, not ODE's, because terms involving derivatives of
$\alpha$ and $\beta$ only enter at higher order in ${1/\ell}$, as one can
verify by direct calculation. Hence, the solutions for $\alpha$ and $\beta$ can
be read off at once, and are:
\bse
\label{eqs:3.5}
\bea
\alpha(x) &=& \frac{16 x}{(1+x^2)^3},
\label{eq:3.5a}\\
\beta(x) &=& 2\,\frac{(3x^4 - 6x^2 - 1)}{x^2(1+x^2)^3}\,g(x) -
2\,\frac{(3x^2-1)}{x(1+x^2)^2}\,g'(x) \nonumber\\
&& + \frac{2}{1+x^2}\,g''(x) + A(x),
\label{eq:3.5b}
\eea
where
\bea
A(x) &=& \alpha''(x) + \frac{1}{x}\,\alpha'(x) - \frac{1}{x^2}\,\alpha(x)
\nonumber\\
&=& \frac{384x(x^2-1)}{(1+x^2)^5}.
\label{eq:3.5c}
\eea
\ese
Similarly, by comparing coefficients in Eq.\ (\ref{eq:2.17a}) we find ODEs
for the functions $g$ and $h$,
\begin{widetext}
\bse
\label{eqs:3.6}
\bea
g''(x) + \frac{1}{x}\,g'(x) - \frac{1}{x^2}\,\cos(2f(x))\,g(x) &=&
-\frac{2}{x}\,\sin(f(x))\,\alpha(x),
\label{eq:3.6a}\\
h''(x) + \frac{1}{x}\,h'(x) - \frac{1}{x^2}\,\cos(2f(x))\,h(x) &=&
-\frac{2}{x}\,\sin(f(x))\,\beta(x) - \frac{1}{x^2}\,\sin(2f(x))\,g^2(x) -
\frac{2}{x}\,\cos(f(x))\,\alpha(x)\,g(x), \nonumber\\
\label{eq:3.6b}
\eea
\ese
\end{widetext}
with $f(x)$ from Eq.\ (\ref{eq:3.6b}).

The ODE (\ref{eq:3.6a}) for $g$ can be solved by standard methods, see Appendix
\ref{app:B}. The physical solution is the one that vanishes for $x\to 0$; it is
proportional to $x$ for $x\gg 1$. We find
\bse
\label{eqs:3.7}
\bea
g(x) = -\frac{4}{3}\,\frac{x[x^2(4+x^2) + 2(1+x^2)\ln(1+x^2)]}{(1+x^2)^2}\ ,
\nonumber\\
\label{eq:3.7a}
\eea
the large-$x$ asymptotic behavior of which is
\bea
g(x\gg 1) = -\frac{4}{3}\,x  -
\frac{16}{3}\,\frac{\ln x}{x}  - \frac{8}{3x}+ O(\frac{\ln x}{x^2}).
\label{eq:3.7b}
\eea
\ese

This determines both the function $\beta(x)$, Eq.\ (\ref{eq:3.5b}), and the
inhomogeneity of the ODE (\ref{eq:3.6b}) for $h(x)$. The latter can again be
solved in terms of tabulated functions, see Appendix \ref{app:B}, but we will
need only the two leading terms for $x\to\infty$. The physical solution is
again the one that vanishes for $x\to 0$, and its large-$x$ asymptotic behavior
is
\bea
h(x\gg 1) = -\frac{32}{9}\,x\ln x + \frac{536}{135}\,x + O(1/x).
\label{eq:3.8}
\eea

Finally, we need to fix the length scale $\ell$. It is determined by the
requirement $\theta(r=R)=0$. We find
\bse
\label{eqs:3.9}
\bea
\ell^2 = \sqrt{\frac{c}{2}}\,R\,\Biggl[1 + \frac{\sqrt{2c}}{R}\,\ln R +
\frac{\delta}{R} + O\left(\frac{\ln R}{R^{3/2}}\right)\Biggr],
\label{eq:3.9a}
\eea
where
\bea
\delta = \sqrt{2c}\,\left[\frac{1}{12}(7-6d/c^2) -
\frac{1}{2}\,\ln(c/2)\right],
\label{eq:3.9b}
\eea
and
\bea
c &=&4/3,
\label{eq:3.9c}\\
d &=& 536/135,
\label{eq:3.9d}
\eea
\ese
are the absolute values of the coefficients of the terms proportional to $x$ in
the large-$x$ expansions of $g(x)$ and $h(x)$, respectively. We see that, for
$R\gg 1$, $\ell$ is indeed proportional to $\sqrt{R}$ , as we had
anticipated above. That is, the characteristic skyrmion length scale
$\ell$ is the geometric mean of the London penetration depth $\lambda$
(recall that we measure all lengths in units of $\lambda$) and the
skyrmion size $R$. We now can also check our requirement $\delta\theta\ll
1$: from Eq.\ (\ref{eq:3.4a}) we see that for $r\ll\ell$,
$\delta\theta(r) \propto 1/R$, while for
$r\gg\ell$, $\delta\theta(r)$ is bounded by a term proportional to $1/R^{1/2}$.
For $R$ large compared to the penetration depth the condition is thus fulfilled
for all $r$. Similarly, $\delta a$ is found to be small compared to
$a_{\infty}$ for all $r$.

\subsection{Energy of a single skyrmion}
\label{subsec:III.C}

By using our perturbative solution in Eq.\ (\ref{eq:2.16}), we are now in a
position to calculate the energy of a single skyrmion to $O(1/R^2)$. It is
convenient to first expand the energy in powers of $1/\ell^2$, and then
determine the $R$-dependence by using Eqs.\ (\ref{eqs:3.9}).

Let us first consider the supercurrent energy $E_{\text{c}}$, i.e., the second
term in Eq.\ (\ref{eq:2.16}). It can be written
\bea
E_{\text{c}}/E_0 = \int_0^R dr\,r\,\left(\delta a(r)\right)^2 &=&
\frac{1}{\ell^6}
\int_0^R dr\,r\,\left(\alpha(r/\ell)\right)^2 \nonumber\\
&& +\  O(1/\ell^6).
\label{eq:3.10}
\eea
Using Eqs.\ (\ref{eq:3.5a}) we find
\be
E_{\text{c}}/E_0 = \frac{32}{5}\,\frac{1}{\ell^4} + O(1/\ell^6).
\label{eq:3.11}
\ee

Now consider the magnetic energy $E_{\text{m}}$, which is the third term in
Eq.\ (\ref{eq:2.16}). It can be written
\bea
E_{\text{m}}/E_0 = \int_0^R dr\,r\,b^2(r),
\label{eq:3.12}
\eea
with
\bse
\label{eqs:3.13}
\bea
b(r) = \frac{1}{r}\,a_{\infty}(r) + a_{\infty}'(r) +
\frac{1}{r}\,\delta a(r) + \delta a'(r),
\label{eq:3.13a}
\eea
the magnetic induction in our reduced units. Notice that in calculating
$a_\infty(r)$, $\theta(r)$ in Eq.\ (\ref{eq:3.1}) needs to be expanded to first
order in $\delta\theta$, as noted earlier. The two leading contributions to
$b^2$ are then
\bea
b^2(r) &=& \frac{16}{\ell^4}\,\frac{1}{(1+x^2)^4} -
\frac{8}{\ell^6}\,\frac{1}{(1+x^2)^2}\,\left[ 2\,\frac{1-x^2}{x(1+x^2)^2}\,g(x)
\right. \nonumber\\
&& \left. + \frac{2g'(x)}{1+x^2} + \frac{1}{x}\,\alpha(x) + \alpha'(x)\right] +
O(1/\ell^8),
\label{eq:3.13b}
\eea
\ese
where $x=r/\ell$. Performing the integral yields
\be
E_{\text{m}}/E_0 = \frac{8}{3}\,\frac{1}{\ell^2} -
\frac{112}{135}\,\frac{1}{\ell^4} + O(1/\ell^6).
\label{eq:3.14}
\ee

Finally, we need to calculate the energy $E_{\text{s}}$ coming from the
gradient terms in the first term in Eq.\ (\ref{eq:2.16}). The expansion of the
two terms in the integrand yields seven integrals that contribute to the
desired order, they are listed in Appendix \ref{app:C}. The result is
\bea
E_{\text{s}}/E_0 &=& 2 + \frac{8}{3}\,\frac{1}{\ell^2} +
\frac{64}{9}\,\frac{\ln\ell}{\ell^4} \nonumber\\
&& + \left(-\,\frac{1,832}{135} - 2c\sqrt{2c}\,\delta + 4c^2
\ln(2/c)\right)\,\frac{1}{\ell^4} \nonumber\\
&& +\, O(1/\ell^6). \nonumber\\
\label{eq:3.15}
\eea

Adding the three contributions, and using Eqs.\ (\ref{eqs:3.9}), we find our
final result for the energy of a skyrmion of radius $R\gg 1$,
\bea
E/E_0 &=& 2 + \frac{8\sqrt{6}}{3}\,\frac{1}{R} - \frac{16}{3}\,\frac{\ln
R}{R^2}
\nonumber\\
&& - \frac{4}{45}\,\left[7 + 30\ln(3/2)\right]\,\frac{1}{R^2} + O(\ln^2
R/R^3). \nonumber\\
\label{eq:3.16}
\eea

Knigavko et al.\cite{Knigavko_Rosenstein_Chen_1999} solved the Eqs.\
(\ref{eqs:2.17}) numerically, and thereby numerically determined the energy,
which they fit to a $1/R$-dependence. Their results are shown in Fig.
\ref{fig:6} together with the analytical result given in Eq.\ (\ref{eq:3.16}).
The perturbative solution up to $O(\ln R/R^2)$ was first given in Ref.\
\onlinecite{Li_Toner_Belitz_2007}. We have also solved the equations
numerically, using spectral methods to convert the boundary value problem to a
set of algebraic equations for the unknown coefficients in an expansion in
Chebyshev polynomials\cite{Boyd_1989}. For the $R$-range shown, and on the
scale of the figure, the result is indistinguishable from the perturbative one.
\begin{figure}[t]
\vskip -0mm
\includegraphics[width=8.5cm]{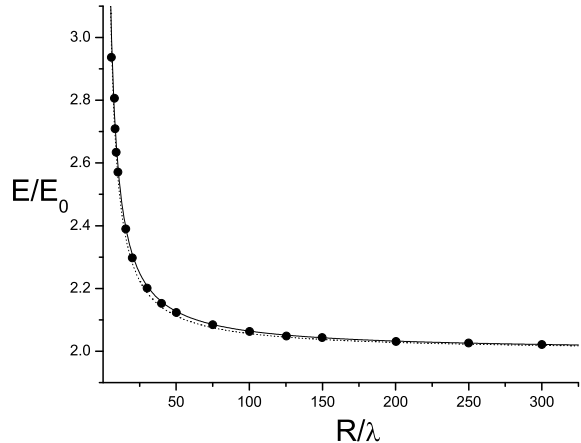}
\vskip -0mm
\caption{Numerical data for the energy per skyrmion per unit length (circles)
  together with the best fit to a pure $1/R$ behavior (dashed line) from Ref.\
  \cite{Knigavko_Rosenstein_Chen_1999}, and the perturbative analytic solution
  given by Eq. (\ref{eq:3.16}) (solid line). A numerical solution using
  spectral methods is indistinguishable from the perturbative one.}
\label{fig:6}
\end{figure}

\section{Observable consequences of the skyrmion energy}
\label{sec:IV}

Our calculation of the skyrmion energy in Sec.\ \ref{sec:III} has been for a
cylindrically symmetric skyrmion. The result shows that each skyrmion will try
to maximize its radius in order to minimize the energy, which leads to a
repulsive interaction between skyrmions whose potential is proportional to
$1/R$. Skyrmions are thus expected to form a lattice structure, as do vortices,
and they will thus {\em not} be cylindrically symmetric, since the lattice is
not. One expects a hexagonal lattice, as in the case of the vortex lattice, and
our treatment involves the same approximation as in the numerical work of Ref.\
\onlinecite{Knigavko_Rosenstein_Chen_1999}; namely, approximating the hexagonal
unit cell by a circle of the same area. We expect this approximation to recover
the correct scaling of the energy, and to reproduce the coefficients of that
scaling to the same accuracy as radius of the circle of the same area
reproduces the distance from the center of a hexagon to the nearest point on
its edge; i.e., $\sqrt{2\sqrt{3}/\pi} -1 \approx 0.05$. We will now proceed to
calculate observable consequences of the dependence of the energy on the radius
of the unit cell. These include the relation $B(H)$ between the magnetic
induction $B$ and the external magnetic field $H$, the elastic properties of
the skyrmion lattice and the resulting phase diagram in the $H$-$T$-plane, and
the $\mu$SR signature of the skyrmion lattice.

\subsection{$B(H)$ for a skyrmion lattice}
\label{subsec:IV.A}

We start by calculating the dependence of the equilibrium lattice constant R on
an external magnetic field $H$. This is done by minimizing the energy per unit
volume, which is the energy per unit length per skyrmion, Eq.\ (\ref{eq:3.16}),
divided by the area per skyrmion, $\pi R^2$, plus a reduction in the energy of
$-2 \Phi_0 H/4\pi$ due to the external field. The latter is obtained from the
last term in Eq.\ (\ref{eq:2.8}) by noting that the magnetic flux $\int dx dy\
(\hat{z}\cdot{\bm b}) = 2 \Phi_0$ for each skyrmion in the lattice. This
negative external field contribution must also be divided by $\pi R^2$ to give
the energy per unit volume. Returning to ordinary units, we thus find a Gibbs
free energy per unit volume
\bea
g(R) = \frac{K}{4\pi^2}\,\left[-\frac{\Delta}{R^2} +
\frac{4\sqrt{6}\lambda}{3R^3} +
O\left(\frac{\lambda^2\ln(R/\lambda)}{R^4}\right)\right]\ , \nonumber\\
\label{eq:4.1}
\eea
where $K = \Phi_0^2/2\pi\lambda^2$, and
\bea
\Delta \equiv 1 - H/\Hcone,
\label{eq:4.2}
\eea
with $\Hcone
\equiv K/2\Phi_0$. For $H < \Hcone$ , we have $\Delta > 0$,  and the free
energy is minimized by $R = \infty$; i.e.,  the skyrmion density is zero.
This is the Meissner phase. For $H > \Hcone$ the free energy is minimized
by
\bea
R = R_0 = 2\sqrt{6}\lambda/\Delta,
\label{eq:4.3}
\eea
and there is a nonzero skyrmion density. We see that $\Hcone$ is indeed the
lower critical field. Note that the equilibrium flux lattice constant $R_0$
diverges as $1/\Delta$, whereas in the case of a vortex lattice it diverges
only logarithmically as $\ln (1/\Delta)$.\cite{Tinkham_1975} For the averaged
magnetic induction $B = 2\Phi_0/\pi R_0^2$ this implies
\bea
B(H) = \frac{1}{3}\,\Hcone\,\Delta^2.
\label{eq:4.4}
\eea
For $H \to \Hcone$ from above, $B(H)$ in the case of a skyrmion lattice thus
vanishes with  zero slope, whereas in the case of a vortex lattice it vanishes
with an infinite slope.\cite{Tinkham_1975} This result, with a slightly
different prefactor, was first obtained from the aforementioned numerical
determination of $E(R)$ in Ref.\ \onlinecite{Knigavko_Rosenstein_Chen_1999}.
Note that the only material parameter that appears in this expression for $B$
is $\Hcone$.

\subsection{Elastic properties of the skyrmion lattice}
\label{subsec:IV.B}

Now we turn to the elastic properties of skyrmion lattice. Let the equilibrium
position of the $i^{\,\text{th}}$ skyrmion line be described by a
two-dimensional lattice vector ${\bm R}_i = (X_i,Y_i)$, and the actual position
by
\bea
{\bm r}_i(z) = (X_i + u_x({\bm R}_i,z), Y_i + u_y({\bm R}_i,z),z),
\label{eq:4.5}
\eea
where ${\bm u} = (u_x,u_y)$ is the two-dimensional displacement vector, and we
use $z$ as the parameter of the skyrmion line. The strain tensor
$u_{\alpha\beta}$ is defined as
\bea
u_{\alpha\beta}({\bm x}) = \frac{1}{2}\,\left(\frac{\partial
u_{\alpha}}{\partial x_{\beta}} + \frac{\partial u_{\beta}}{\partial
x_{\alpha}}\right).
\label{eq:4.6}
\eea
For a hexagonal lattice of lines parallel to the $z$-axis, the elastic
Hamiltonian is\cite{Landau_Lifshitz_VII_1986}
\bea
H_{\text{el}} &=& \frac{1}{2}\int d{\bm x}\ \bigl[2\mu\,
(u_{\alpha\beta}({\bm x})u_{\alpha\beta}({\bm x})) +
\lambda_{\text{L}}\,(u_{\alpha\alpha}({\bm x}))^2 \nonumber\\ &&\hskip
40pt + K_{\text{tilt}}\vert\partial_z{\bm u}({\bm x})\vert^2 \bigr].
\label{eq:4.7}
\eea
Here summation over repeated indices is implied. $\mu$, $\lambda_{\text{L}}$,
and $K_{\text{tilt}}$ are the shear, bulk, and tilt moduli, respectively, of
the lattice, and we now need to determined these elastic constants.

The combination $\mu + \lambda_{\text{L}}$ can be obtained by considering the
energy change of the system upon a dilation of the lattice. Let $R$ change from
$R_0$ to $R_0 (1 + \epsilon)$, with a dilation factor $\epsilon \ll 1$. Such a
dilation corresponds to a displacement field ${\bm u}({\bm x}) = \epsilon\,{\bm
x}_{\perp}$, where ${\bm x}_{\perp}$ is the projection of ${\bm x}$
perpendicular to the $z$-axis.\cite{Landau_Lifshitz_VII_1986} The strain tensor
is thus $u_{\alpha\beta} = \epsilon\,\delta_{\alpha\beta}$. Inserting this in
the elastic Hamiltonian, Eq.\ (\ref{eq:4.7}), yields the energy per unit volume
for the dilation,
\bse
\label{eqs:4.8}
\be
E_{\text{dil}}/V = 2(\mu + \lambda_{\text{L}})\,\epsilon^2.
\label{eq:4.8a}
\ee
This should be compared with the energy as given by Eq.\ (\ref{eq:4.1}),
\bea
E_{\text{dil}}/V &=& g(R_0(1+\epsilon)) - g(R_0) =
\frac{1}{2}\,\left(\frac{\partial^2 g}{\partial R^2}\right)_{R_0} (\epsilon
R_0)^2 \nonumber\\
&=& \frac{K\Delta^3}{96\pi^2\lambda^2}\,\epsilon^2.
\label{eq:4.8b}
\eea
\ese
Comparing Eqs.\ (\ref{eq:4.8a}) and (\ref{eq:4.8b}) yields
\be
\mu + \lambda_{\text{L}} = K\,\Delta^3/192\pi^2\lambda^2.
\label{eq:4.9}
\ee

To obtain $\mu$ (or $\lambda_{\text{L}}$) separately, we should consider shear
deformations, which change the shape, but not the area, of the unit cell. Since
we have already approximated the hexagonal unit cell by a circle, this is
difficult to do, and we resort to the following heuristic method, which will
give the correct scaling of $\mu$ with $\Delta$ (but not the correct
prefactors). To this end we observe that our result for the Gibbs free energy,
Eq.\ (\ref{eq:4.1}), is of the same form we would have obtained if the
skyrmions interacted via a pair potential $U(r)$ that for distances $r\alt R_0$
is of order $K\lambda/r$, and for larger distances falls off sufficiently
rapidly that only nearest-neighbor interactions need to be considered. Treating
the skyrmion lattice as if such an ``equivalent potential'' were the origin of
the skyrmion energy allows us to calculate the shear modulus as follows:

If the lattice is subjected to a uniform $x-y$ shear - i.e., a displacement
field ${\bm u}({\bm x}) = 2 \epsilon y \hat{\bm x}$ - for which $u_{xy} =
u_{yx} = \epsilon$, and all other components of $u_{\alpha\beta} = 0$, the
elastic energy, Eq.\ (\ref{eq:4.7}) predicts an elastic energy per unit volume
of
\bea
E/V = 2\mu \epsilon^2.
\label{eq:4.10}
\eea

Such a shear skews each fundamental triangle of the skyrmion lattice by
displacing the top (or bottom, for the downward-pointing triangles) to the
right (or the left, for downward-pointing triangles) by an amount of order
$\epsilon R_0$, where $R_0$ is the skyrmion lattice spacing found earlier, Eq.\
(\ref{eq:4.3}) (see Fig.\ \ref{fig:7}).
\begin{figure}[t,h]
\includegraphics[width=5cm]{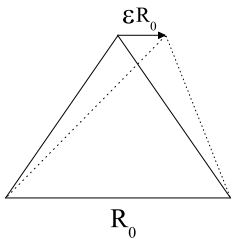}
\vskip -0mm
\caption{Shearing of the skyrmion lattice results in a change in the distance
 between skyrmion centers, and hence in their effective interaction. See the
 text for additional information.}
\label{fig:7}
\end{figure}
This shortens the length of one bond of the triangle by an amount of order
$\epsilon R_0$, and increases the opposite bond's length by the same amount.
Hence, the linear in $\epsilon$ change in the ``equivalent potentials'' of
these two bonds cancels, and the total change $(\Delta E/{\text{triangle}})$ in
the energy per unit length of fundamental triangle, per triangle, is given by:
\bea
{ \Delta E\over \text{triangle}} = U''(R_0) (\epsilon R_0)^2 \times O(1)
\label{eq:4.11}
\eea
where the $O(1)$ factor includes both geometrical factors (e.g., sines
and cosines), and counting factors (e.g., to avoid multiple counting of
each triangle). If we take $U(r) = K\lambda/r$ as suggested above, we
have
\bea
U''(R_0) = {K\lambda\over R_0^3}\times O(1).
\label{eq:4.12}
\eea
Inserting this into Eq.\ (\ref{eq:4.11}) gives
\bea
{ \Delta E\over \text{triangle}} = {K\lambda\over R_0} \epsilon^2 \times O(1).
\label{eq:4.13}
\eea
This is the change in energy per unit cell.
To get the energy per unit volume, we must divide by the unit cell area,
which is $\pi R_0^2$. Doing so gives
\bea
{ \Delta E\over V} = {K\lambda\over R_0^3} \epsilon^2 \times O(1).
\label{eq:4.14}
\eea
Comparing this with Eq.\ (\ref{eq:4.10}) then determines $\mu$:
\bea
\mu = {K\lambda\over R_0^3}  \times O(1).
\label{eq:4.15}
\eea
Using Eq.\ (\ref{eq:4.3}) for $R_0$ then leads to our final result for $\mu$:
\bse
\label{eqs:4.16}
\bea
\mu = \frac{K\Delta^3}{\lambda^2} \times O(1).
\label{eq:4.16a}
\eea
 From Eq.\ (\ref{eq:4.9}) we see that the bulk modulus or Lam{\`e} coefficient
is given given by the same expression,
\bea
\lambda_{\text{L}} =  \frac{K\Delta^3}{\lambda^2} \times O(1).
\label{eq:4.16b}
\eea
\ese

We now turn to the tilt modulus $K_{\text{tilt}}$. This can be obtained by
considering a uniform tilt of the axes of the skyrmions away from the $z$-axis,
i.e., away from the direction of the external magnetic field $H$, by an angle
$\vartheta\ll 1$. For small $\vartheta$, $\vartheta = \vert\partial{\bm
u}/\partial z\vert$. Therefore, the tilt energy in Eq.\ (\ref{eq:4.7}) is
identical with the change of the ${\bm B}\cdot{\bm H}$ term in Eq.\
(\ref{eq:2.8}). This contribution to the energy is, per unit length and in
ordinary units, given by $-\Phi_0\,H\cos\theta/2\pi$, and its change due to
tilting is $\Phi_0\,H(1 - \cos\vartheta)/2\pi \approx
\Phi_0\,H\,\vartheta^2/4\pi = \Phi_0\,H\,\vert\partial_z{\bm u}\vert^2/4\pi$.
Dividing this result by the unit cell area $\pi R_0^2$, using Eq.\
(\ref{eq:4.3}) for $R_0$, and identifying the result with the tilt term in the
elastic Hamiltonian, Eq.\ (\ref{eq:4.7}), yields $K_{\text{tilt}}$ in the
vicinity of $\Hcone$,
\be
K_{\text{tilt}} = \frac{1}{12\pi}\,\Hcone^2\, \Delta^2.
\label{eq:4.17}
\ee

We now are in a position to calculate the mean-square positional fluctuations
$\langle\vert{\bm u}({\bm x})\vert^2\rangle$. Taking the Fourier transform of
Eq.\ (\ref{eq:4.7}), and using the equipartition theorem, yields
\bse
\label{eqs:4.18}
\be
\langle\vert{\bm u}({\bm x})\vert^2\rangle_{\text{T}} = \frac{k_{\text{B}}T}{V}
\sum_{\bm q\in \text{BZ}} \frac{1}{\mu \,q_{\perp }^2 + K_{\text{tilt}}\,q_z^2}
\label{eq:4.18a}
\ee
for the transverse fluctuations, and
\be
\langle\vert{\bm u}({\bm x})\vert^2\rangle_{\text{L}} = \frac{k_{\text{B}}T}{V}
\sum_{\bm q\in \text{BZ}} \frac{1}{(2\mu + \lambda_{\text{L}}) \,q_{\perp }^2 +
K_{\text{tilt}}\,q_z^2}
\label{eq:4.18b}
\ee
\ese
for the longitudinal ones. Here ${\bm q}_{\perp}$ and $q_z$ are the projections
of the wave vector ${\bm q}$ orthogonal to and along the $z$-direction,
respectively. The Brillouin zone BZ of the skyrmion lattice is a hexagon (which
we have approximated by a circle) of edge length $O(1)/R_0$ in the plane
perpendicular to the $z$-axis, and extends infinitely in the $z$-direction.

Since $\mu$ and $\lambda_{\text{L}}$ are the same apart from a prefactor of
$O(1)$ which we have not determined, see Eqs.\ (\ref{eqs:4.16}), the same is
true for the transverse and longitudinal contributions to the fluctuations, and
it suffices to consider the former. Performing the integral over $q_z$ yields
\bea
\langle\vert{\bm u}({\bm x})\vert^2\rangle &=& \langle\vert{\bm u}({\bm
x})\vert^2\rangle_{\text{L}} + \langle\vert{\bm u}({\bm
x})\vert^2\rangle_{\text{T}} \nonumber\\
&\propto& \langle\vert{\bm u}({\bm x})\vert^2\rangle_{\text{T}} =
\frac{k_{\text{B}}T}{\sqrt{\mu K_{\text{tilt}}}} \int_{\text{BZ}} \frac{d^2
q_{\perp}}{8\pi^2}\,\frac{1}{q_{\perp}}\ . \nonumber\\
\label{eq:4.19}
\eea
The remaining integral over the perpendicular part of the Brillouin zone is
proportional to $1/R_0$, and using Eqs.\ (\ref{eqs:4.16}) and (\ref{eq:4.3}) we
obtain
\be
\langle\vert{\bm u}({\bm x})\vert^2\rangle =
\frac{k_{\text{B}}T}{\lambda\Hcone^2}\,\frac{1}{\Delta^{3/2}} \times O(1).
\label{eq:4.20}
\ee
Using Eq.\ (\ref{eq:4.3}) again we see that, near $\Hcone$, $\langle\vert{\bm
u}({\bm x})\vert^2\rangle \propto R_0^{3/2} \ll R_0^2$. That is, in this regime
the positional fluctuations are small compared to the lattice constant, which
tells us that the lattice will be stable against melting. To elaborate on this,
let us consider the Lindemann criterion for melting, which states that the
lattice will melt when the ratio $\Gamma_{\text{L}} = \langle\vert{\bm u}({\bm
x})\vert^2\rangle/R_0^2$ exceeds a critical value $\Gamma_{\text{c}} = O(1)$.
In our case,
\be
\Gamma_{\text{L}} =
\frac{k_{\text{B}}T}{\Hcone^2\,\lambda^{5/2}}\,\Delta^{1/2}\,\times O(1).
\label{eq:4.21}
\ee
As $H \to \Hcone$, $\Delta \to 0$, and the Lindemann ratio vanishes. Hence, the
skyrmion lattice does not melt at any temperature for $H$ close to $\Hcone$.

We finally determine the shape of the melting curve $H_{\text{m}}(T)$ near the
superconducting transition temperature $T_{\text{c}}$. Since, in mean field
theory, $\Hcone \propto (T_{\text{c}}-T)$, and $\lambda \propto
1/\sqrt{T_{\text{c}}-T}$,\cite{Tinkham_1975} we find from Eq. (\ref{eq:4.21})
by putting $\Gamma_{\text{L}} = \text{const.} = O(1)$,
\be
H_{\text{m}} - \Hcone \propto
(T_{\text{c}}-T)^{5/2}.
\label{eq:4.22}
\ee

The resulting phase diagram is shown schematically in Fig.\ \ref{fig:8}.
Comparing with Fig.\ \ref{fig:1} we see the qualitative difference between the
vortex and skyrmion flux lattices: whereas the vortex lattice always melts near
$\Hcone$, the skyrmion lattice melts nowhere near $\Hcone$. This is a direct
consequence of the long-ranged interaction between skyrmions, as opposed to the
screened Coulomb interaction between vortices.
\begin{figure}[t,h]
\includegraphics[width=5cm]{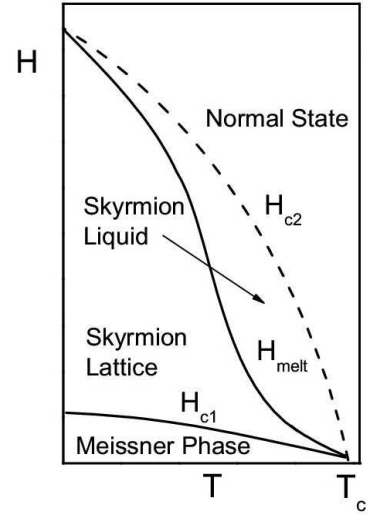}
\vskip -0mm
\caption{External field ($H$) vs. temperature ($T$) phase diagram for
  skyrmion flux lattices. In contrast to the vortex case, see Fig.
\ref{fig:1},
  there is a direct transition from the skyrmion flux lattice to the Meissner
  phase. The theory predicts the shape of the melting curve only close to
  $T_{\text{c}}$, see Eq.\ (\ref{eq:4.22}); the rest of the curve is an educated
  guess.}
\label{fig:8}
\end{figure}

\subsection{$\mu$SR signature of a skyrmion flux lattice}
\label{subsec:IV.B}

Muon spin rotation ($\mu SR$) is a powerful tool which has been extensively
applied to study the vortex state in type-II superconductors.
\cite{Sonier_Brewer_Kiefl_2000, Schenck_1986} A crucial quantity in this type
of experiment is the $\mu$SR line shape $n(B)$, which is the probability
density that a muon experiences a local magnetic induction $B$ and precesses at
the Larmor frequency that corresponds to $B$. It is defined as
\bea
n(B) \equiv \langle \delta(B({\bm x}) - B)\rangle,
\label{eq:4.23}
\eea
where $B({\bm x})$ is the magnitude of the local magnetic induction, and
$\langle \ldots \rangle$ denotes the spatial average over a flux lattice unit
cell.

To predict the $\mu$SR line shape for a skyrmion flux lattice near $\Hcone$ it
is sufficient, for large $R_0$,  to  use only the lowest solution for the
magnetic induction obtained in Sec.\ \ref{subsec:III.A}. Inserting Eqs.\
(\ref{eqs:3.3}) into Eq.\ (\ref{eq:3.1}), we find for the magnetic induction in
reduced units
\bea
b(r) = -\frac{4\ell^2}{(r^2 + \ell^2)^2}.
\label{eq:4.24}
\eea
Restoring physical units then gives
\bea
B(r) = \frac{\Hcone \lambda^2}{2}\frac{\ell^2}{(r^2 + \ell^2)^2},
\label{eq:4.25}
\eea
where we've dropped the minus sign since only the magnitude of $B$ can be
detected in $\mu$SR measurements.

From Eq.\ (\ref{eq:4.23}) we then find, for $H$ near $\Hcone$, where our theory
is valid,
\bea
n(B) = {1\over 24 \sqrt{2}} \left({\Hcone\Delta\over B}\right)^{{3\over 2}}{1
\over \Hcone} \hskip 8pt (\text{skyrmions}).
\label{eq:4.26}
\eea

Of course, $n(B)$ is only non-zero for those values of $B$ that actually occur
inside the unit cell of the skyrmion lattice. From Eq.\ (\ref{eq:4.25}), we see
that the maximum value of $B$ will occur at the center of the unit cell
($r=0$), which gives
\bse
\label{eqs:4.27}
\bea
\vert B\vert_{\text{max}} = \vert B(r=0)\vert = \frac{\Hcone
\lambda^2}{2\ell^2} = \frac{\Hcone \Delta}{8},
\label{eq:4.27a}
\eea
The minimum value of $B$ occurs at the edge of the unit cell (i.e., $r=R$),
where Eq.\ (\ref{eq:4.25}) gives
\bea
\vert B\vert_{\text{min}} = \vert B(r=R)\vert = \frac{\Hcone
\lambda^2\ell^2}{2R^4} = \frac{\Hcone \Delta^3}{288}.
\label{eq:4.27b}
\eea
\ese
In the second equalities in Eqs.\ (\ref{eqs:4.27}) we have used Eqs.\
(\ref{eqs:3.9}) and (\ref{eq:4.3}) to express $\ell$ in terms of $R$ and $R$ in
terms of $\Delta$, respectively.

To summarize: the prediction of our cylindrical approximation for $n(B)$ is
that the simple power law  Eq.\ (\ref{eq:4.26}) holds for $B_{\text{min}} < B <
B_{\text{max}}$. For $B < B_{\text{min}}$ or $B > B_{\text{max}}$, $n(B) = 0$.

Since the above results were derived in the cylindrical approximation, we
expect the numerical coefficients in Eqs. (\ref{eqs:4.27}) to be off by the
approximately $5\%$ mentioned in the opening paragraph of Sec.\ \ref{sec:IV}
throughout most of the range $B_{\text{min}} < B < B_{\text{max}}$. When $B$
gets close to $B_{\text{min}}$, however, we expect more radical departures from
the cylindrical approximation. This is because contours of constant $B$ near
the edge of the hexagonal unit cell will, for $B$ within $5\%$ or so of
$B_{\text{min}}$ or so, start intersecting the unit cell boundary, leading to
van Hove-like singularities in $n(B)$. Such subtleties cannot be captured
within the cylindrical approximation. Note, however, that they only occur over
a very small range of $B$; for the remainder of the large window
$B_{\text{min}} < B < B_{\text{max}}$ (which spans three decades even for
$\Delta $ as big as $0.2$), Eq.\ (\ref{eq:4.26}) holds, up to the
aforementioned $5\%$ numerical error in its overall coefficient.

To compare this result with the corresponding one for a vortex flux lattice, we
recall that in that case $B(r)$ is given by a modified Bessel function which
for distances $r \gg \lambda$ takes the form
\bea
B(r) \propto \frac{1}{\sqrt{r/\lambda}}\,e^{-r/\lambda}.
\label{eq:4.28}
\eea
For small $B$, we then find from Eq.\ (\ref{eq:4.23})
\bea
n(B) \propto \frac{\ln(1/B)}{B} \hskip 30pt (\text{vortices}).
\label{eq:4.29}
\eea

We see that the $\mu$SR line shape is qualitatively different in the two cases,
due to the long-range nature of $B(r)$ in the skyrmion case versus the
exponential decay in the vortex case.

\section{Conclusion}
\label{sec:V}

In summary, we have considered properties of a flux lattice formed by the
topological excitations commonly referred to as skyrmions, rather than by
ordinary vortices. For strongly type-II materials in the $\beta$-phase,
skyrmions are more stable than vortices.\cite{Knigavko_Rosenstein_Chen_1999} We
have presented an analytical calculation of the energy of a cylindrically
symmetric skyrmion of radius $R$ up to $O(1/R^2)$ in an expansion in powers of
$1/R$. This provides excellent agreement with numerical solutions of the
skyrmion equations. The interaction between skyrmions is long-ranged, falling
off only as the inverse distance, in contrast to the exponentially decaying
interaction between vortices. As a result, the elastic properties of a skyrmion
flux lattice are very different from those of a vortex flux lattice, which
leads to qualitatively different melting curves for the two systems. The phase
diagram thus provides a smoking gun for the presence of skyrmions. In addition,
the $\mu$SR line width for skyrmions is qualitatively different from the vortex
case.

We finally mention two limitations of our discussion. First, we have restricted
ourselves to a discussion of a particular $p$-wave ground state, namely, the
non-unitary state sometimes referred to as the $\beta$-phase. This state breaks
time-reversal symmetry and the recently reported absence of experimental
evidence for the latter in Sr$_2$RuO$_4$\cite{Bjornsson_et_al_2005} suggests to
also consider other possible $p$-wave states and their topological excitations,
in analogy to the rich phenomenology in Helium 3.\cite{Salomaa_Volovik_1987}
Second, in a real crystalline material, crystal-field effects will invalidate
our isotropic model at very long distances, and cause the skyrmion interaction
to fall off exponentially. This is the same effect that makes, for instance,
the isotropic Heisenberg model of ferromagnetism inapplicable at very long
distances and gives the ferromagnetic magnons a small mass. It should be
emphasized that this is usually an extremely weak effect that is also material
dependent. Once $p$-wave superconductivity has been firmly established in a
particular material, this point needs to be revisited in order to determine the
energy scales on which the above analysis is valid.

\acknowledgments

We thank Tom Devereaux for suggesting the discussion of $\mu$SR as a possible
probe for skyrmion lattices, and Hartmut Monien for a discussion on defects in
Helium 3. Part of the work was performed at the Aspen Center for Physics. This
work was supported by the NSF under grant No. DMR-05-29966.

\appendix\ \

\section{Properties of orthogonal unit vectors}
\label{app:A}

Let $\hat{\bm n}$ and $\hat{\bm m}$ be orthogonal real unit vectors, and
$\hat{\bm l} = \hat{\bm n}\times\hat{\bm m}$. Then the normalization condition
${\hat n}_i {\hat n}_i = {\hat m}_i {\hat m}_i = 1$ and the orthogonality
condition ${\hat n}_i {\hat m}_i = 0$ imply
\bse
\label{eqs:A.1}
\bea
{\hat n}_j \p_i {\hat n}_j &=& {\hat m}_j \p_i {\hat m}_j = 0,
\label{eq:A.1a}\\
{\hat n}_j \p_i {\hat m}_j &=& -{\hat m}_j \p_i {\hat n}_j.
\label{eq:A.1b}
\eea
\ese
With these relations it is straightforward to show that
\bea
\p_i {\hat n}_j \p_i {\hat n}_j + \p_i {\hat m}_j \p_i {\hat m}_j = 2({\hat
n}_j\p_i {\hat m}_j)({\hat n}_k \p_i {\hat m}_k) + \p_i {\hat l}_j \p_i {\hat
l}_j. \nonumber\\
\label{eq:A.2}
\eea
Finally, in regions where ${\hat l}({\bm x})$ is differentiable the Mermin-Ho
relation\cite{Mermin_Ho_1976} holds,
\bea
{\hat {\bm l}}\cdot(\p_i{\hat {\bm l}}  \times \p_j{\hat {\bm l}}) = \p_i
{\hat {\bm n}}\cdot \p_j {\hat {\bm m}} - \p_i {\hat {\bm m}}\cdot \p_j
{\hat {\bm n}}.
\label{eq:A.3}
\eea

\section{Solutions of the ODE$s$ for $g$ and $h$}
\label{app:B}

The functions $g$ and $h$ in Sec.\ \ref{subsec:III.B} both satisfy an ODE of
the form (see Eqs.\ (\ref{eqs:3.6}))
\bea
F''(x) + \frac{1}{x}\,F'(x) - \frac{(x^4-6x^2+1)}{x^2(1+x^2)^2}\,F(x) =
q(x),
\label{eq:B.1}
\eea
with an inhomogeneity $q$ given by the right-hand side of Eq.\ (\ref{eq:3.6a})
or (\ref{eq:3.6b}), respectively. It is easy to check that the corresponding
homogeneous equation, obtained from Eq.\ (\ref{eq:B.1}) by putting $q(x)\equiv
0$, is solved by
\be
F_{\text{h}}(x) = x/(1+x^2).
\label{eq:B.2}
\ee
(This is the solution that vanishes as $x\to 0$. The second solution diverges
in this limit.) Now write $F(x) = F_{\text{h}}(x)\,G(x)$, and let $y(x) =
G'(x)$. Then $y$ is found to obey the elementary first-order ODE
\bse
\label{eqs:B.3}
\be
y'(x) + p(x) y(x) = q(x)/F_{\text{h}}(x),
\label{eq:B.3a}
\ee
with
\be
p(x) = [2F_{\text{h}}'(x) + F_{\text{h}}(x)/x]/F_{\text{h}}(x).
\label{eq:B.3b}
\ee
\ese
The solution is
\be
y(x) = e^{-\int dx\,p}\left[C_1 + \int dx\,q\,e^{\int dx\,p}\right],
\label{eq:B.4}
\ee
with $C_1$ an integration constant. A second integration yields $G(x)$, and
hence $F(x)$ in terms of two integration constants. The latter can be
determined by requiring that for small $x$ the solution coincides with the
asymptotic solution that vanishes as $x\to 0$. By using a power-law ansatz for
$g$ and $h$ in Eqs.\ (\ref{eqs:3.6}) we find $g(x\to 0) = -8x^3 + O(x^4$, and
$h(x\to 0) = 256 x^3 + O(x^4)$, which suffices to fix the integration
constants. For $g(x)$ we find the expression given in Eq.\ (\ref{eq:3.7a}). For
$h(x)$ we obtain
\begin{widetext}
\bea
h(x) &=& \frac{1}{(270 x (1 + x^2)^4)}\,\left\{592 +
   2 x^2 \left[8 (-1,119 + 90 x^2 + 286 x^4 + 240 x^6 + 30 x^8) +
      10,320 (1 + x^2)^3\right]\right.
\nonumber\\
      &&  + 2,296 (-1 + x^2) (1 + x^2)^4 +
   4 x^2 (1 + x^2)^3 +1,704 \ln x
\nonumber\\
&& + 32 \ln(1 + x^2) \left[-30 + 142 x^2 + 276 x^4 + 171 x^6 + 52 x^8 -
      15 x^{10} - 15 (3 + x^2) (x + x^3)^2 \ln(1 + x^2)\right]
      \nonumber\\
&& \left. - 1,920 x^2 (1 + x^2)^3 \text{Li}_2(-x^2)\right\},
\label{eq:B.5}
\eea
\end{widetext}
with Li the polylogarithm function. The asymptotic behavior for large $x$ is
given by Eq.\ (\ref{eq:3.8}).

\section{Contributions to $E_{\text{s}}$}
\label{app:C}

By expanding the integrand of the first term in Eq.\ (\ref{eq:2.16}), we can
express the energy $E_{\text{s}}$ to $O(1/R^2)$ in terms of seven integrals,
\bea
E_{\text{s}}/E_0 = \sum_{i=1}^7 I_i + O(1/\ell^6),
\label{eq:C.1}
\eea
with
\bse
\label{eqs:C.2}
\bea
I_1 &=& 4 \int_0^{R/\ell} dx\, \frac{x}{(1+x^2)^2},
\label{eq:C.2a}\\
I_2 &=& \frac{2}{\ell^2} \int_0^{R/\ell} dx\,
\frac{1}{1+x^2}\,\left(\frac{(x^2-1)}{x^2+1}\,g(x) - x g'(x)\right),
\nonumber\\
\label{eq:C.2b}\\
I_3 &=& \frac{1}{2\ell^4} \int_0^{R/\ell} dx\, \left[x\left(g'(x)\right)^2 +
\frac{(x^4-6x^2+1)}{x(1+x^2)^2}\,g^2(x)\right],\nonumber\\
\label{eq:C.2c}
\eea
\bea
I_4 &=& \frac{2}{\ell^4} \int_0^{R/\ell} dx\,
\frac{1}{1+x^2}\,\left(\frac{(x^2-1)}{x^2+1}\,h(x) - x h'(x)\right),
\nonumber\\
\label{eq:C.2d}\\
I_5 &=& \frac{1}{\ell^6} \int_0^{R/\ell} dx\, \Bigl(x g'(x) h'(x) \nonumber\\
&& \hskip 50pt + \frac{(x^4-6x^2+1)}{x(1+x^2)^2}\,g(x) h(x)\Bigr),
\label{eq:C.2e}\\
I_6 &=& -\frac{4}{3\ell^6} \int_0^{R/\ell} dx\,
\frac{(x^2-1)}{(1+x^2)^2}\,g^3(x),
\label{eq:C.2f}\\
I_7 &=& -\frac{1}{6\ell^8} \int_0^{R/\ell} dx\,
\frac{(x^2-1)^2}{x(1+x^2)^2}\,g^4(x).
\label{eq:C.2g}
\eea
\ese
Evaluating the integrals to $O(1/\ell^4)$ yields Eq. (\ref{eq:3.15}).


\begin{thebibliography}{35}
\bibitem[{sky()}]{skyrmion_footnote}
\bibinfo{note}{Defects of this general type are known under various names in
  different contexts, and the action for the defects studied here differs from
  the one considered by Skyrme. We follow a recent trend to refer to all
  defects of this type as skyrmions.}
\bibitem[{n=1()}]{n=1_footnote}
\bibinfo{note}{More generally, $\phi$ is an element of the circle or one-sphere
  $S_1$, and hence $\oint_{\cal C} d{\bm\ell}\cdot{\bm\nabla}\phi({\bm x})=2\pi
  n$ with $n$ an integer. $n$ is a topological invariant that characterizes the
  singularity (known as a vortex), and the number of flux quanta that penetrate
  the vortex is $N = \vert n \vert$. The vortex with $n=1$ has the lowest
  energy within this family of solutions (apart from the trivial ``non-vortex''
  with $n=0$).}
\bibitem[{Q=1()}]{Q=1_footnote}
\bibinfo{note}{More generally, $\int dx\,dy\ \epsilon_{ij}\,{\hat{\bm l}} \cdot
  (\p_{i}{\hat{\bm l}}\times \p_j {\hat{\bm l}}) = 8\pi Q$, with $Q$ an
  integer. $Q$ is a topological invariant that characterizes the defect (known
  as a skyrmion), and the number of flux quanta that penetrate the skyrmion is
  $N = 2Q$.\cite{Knigavko_Rosenstein_Chen_1999} The skyrmion with $Q=1$ has the
  lowest energy within this family of solutions.}
\bibitem[{exp()}]{expansion_footnote}
\bibinfo{note}{We emphasize that we do {\em not} require $\vert\delta\theta
  (r)\vert \ll \vert\theta_{\infty}(r)\vert$, as this requirement is neither
  necessary nor desirable. Rather, we expand, for instance,
  $\sin(\theta_{\infty} + \delta\theta) = \sin\theta_{\infty} \cos\delta\theta
  + \cos\theta_{\infty} \sin\delta\theta = \cos\theta_{\infty}\,\delta\theta +
  O(\delta\theta^2)$, which is valid for {\em all} $\delta\theta \ll 1$, {\em
  not} just for those that satisfy $\vert\delta\theta (r)\vert \ll
  \vert\theta_{\infty}(r)\vert$.}
\end{thebibliography}

\end{document}